\documentclass[12pt]{article}
\usepackage{amsfonts}
\usepackage{amssymb}
\usepackage{graphicx}
\usepackage{amsmath}
\newcommand{\email}[1]{\vspace*{5pt} {E-mail: \tt{#1}}}

\newcommand{\Fc}{\mathcal{F}}

\begin{document}

\title{\textbf{Localization of the SFT inspired Nonlocal Linear Models and Exact
Solutions}}

\author{\textit{Sergey  Yu. Vernov}\\
{}\\
Skobeltsyn Institute of Nuclear Physics, \\
Moscow State University,\\ Vorobyevy Gory, GSP-1, Moscow, 119991, Russia,\\
\small{\email{svernov@theory.sinp.msu.ru}}}

\date{ }

\maketitle

\begin{abstract}
A general class of gravitational models driven by a nonlocal scalar
field with a linear or quadratic potential is considered. We study the
action with an arbitrary analytic function $\Fc(\Box_g)$, which has
both simple and double roots. The way of localization of nonlocal
Einstein equations is generalized on models with linear potentials.
Exact solutions in the Friedmann--Robertson--Walker and Bianchi I
metrics are presented.
\end{abstract}


\section{Introduction}

Recently a wide class of nonlocal cosmological models based on the
string field theory (SFT) (for details see reviews~\cite{review-sft})
and the $p$-adic string theory \cite{padic} emerges and attracts a lot
of attention \cite{IA1}--\cite{Koshelev2009}. Due to the presence of
phantom excitations nonlocal models are of interest for the present
cosmology. Generally speaking, models that violate the null energy
condition (NEC) have ghosts, and therefore  are unstable and physically
unacceptable. Phantom fields look harmful to the theory and a local
model with a phantom scalar field is not acceptable from the general
point of view. Models with higher derivative terms produce well-known
problems with quantum instability~\cite{AV-NEC,RAS}.   A idea that
could solve the problems is  that terms with high order derivatives can
be treated as corrections valued only at small energies below the
physical cut-off~\cite{SW,Creminelli0812}. This approach implies the
possibility to construct a UV completion of the theory and requires
detailed analysis.

Note that the possibility of the existence of dark energy with
$w_{\mathrm{DE}}<-1$ is not excluded experimentally. Contemporary
cosmological observational data~\cite{data} strongly support that the
present Universe exhibits an accelerated expansion providing thereby an
evidence for a dominating dark energy component with the state
parameter
\begin{equation}
w_{\mathrm{DE}}={} -1.0\pm 0.2.
\end{equation}

The present cosmological observations do not exclude an evolving
parameter~$w_{\mathrm{DE}}$. Moreover, the recent analysis of the
observation data indicates that the varying in time dark energy with
the state parameter $w_{\mathrm{DE}}
$, which crosses the cosmological
constant barrier, gives a better fit than a cosmological
constant~\cite{ZhangGui} (see also~\cite{Starobinsky09,Quinmodrev1} and
references therein).

To obtain a stable model with $w_{\mathrm{DE}}<-1$ one should construct
the effective theory with the NEC violation from the fundamental
theory, which is stable and admits quantization.   From this point of
view the NEC violation might be a property of a model that approximates
the fundamental theory and describes some particular features of the
fundamental theory. With the lack of quantum gravity, we can just trust
string theory or deal with an effective theory admitting  the UV
completion.
 It can be considered as a hint towards the SFT inspired
cosmological models (details about the string cosmology see in
reviews~\cite{string-cosmo}). Note also, that not only the string
inspired cosmological  models obey nonlocality~\cite{nonlocal}.

In the flat space-time nonlocal equations are actively investigated as
well~\cite{PaisU,STinspired,Yang,AJK}. Note that differential equations
of infinite order were began to study long time ago~\cite{davis,carmi}.

The purpose of this paper is to study gravitational models with a
 nonlocal scalar field. We consider a
 general form of a nonlocal action for the scalar field with a quadratic or linear
  potential, keeping the main ingredient, the analytic function $\Fc(\Box_g)$, which
in fact produces the nonlocality in question, almost unrestricted.

 The possible way to find solutions of the Einstein equations
with a quadratic potential of the nonlocal scalar field, is to reduce
them to a system of Einstein equations describing many non-interacting
 local scalar fields~\cite{Koshelev07,AJV0711} (see
also~\cite{KV,Vernov2010}). Some of the obtained local scalar fields
are normal and other of them are phantom ones. In this paper we
generalize the algorithm of localization, proposed
in~\cite{AJV0711,Vernov2010}, on the case of a linear potential. Note
that the way of localization in the case of a linear potential and in
the case of a quadratic potential with a linear term, considered
in~\cite{MN}, are different.

The paper is organized as follows. In Section~2 we describe  nonlocal
cosmological models. In Section 3 we propose the algorithm to find
particular solutions of the nonlocal Einstein equations, solving only
local ones, and prove the self-consistence of it. Any solution for the
obtained system of differential equations is a particular solution for
the initial nonlocal Einstein equations. Exact solutions in the
Friedmann--Robertson--Walker and Bianchi I metrics are presented in
Section~4. In Section~5 we summarize the obtained results and propose
directions for further investigations.


\section{Model setup}

The four-dimensional action with a quadratic or linear potential,
motivated by the string field theory, has been studied
in~\cite{Koshelev07,AJV0701,AJV0711,Vernov:2008hd,MN,KV,Vernov2010,Koshelev2009}.
Such a model appears as a linearization of the SFT inspired model in
the neighborhood of an extremum of the potential (see~\cite{KV} for
details). For linear models, solving the nonlocal equations using the
technique, proposed in~\cite{AJV0711}, is completely equivalent to
solving the equations using the diffusion-like partial differential
equations~\cite{MN}.  By linearizing a nonlinear model about a
particular field value, one is able to specify initial data for
nonlinear models, which he then evolves into the full nonlinear regime
using the diffusion-like equation~\cite{MN}.

In this paper we study nonlocal cosmological models with a quadratic
potential, in other words, a linear nonlocal model, which can be
described by the following action:
\begin{equation}
S=\int d^4x\sqrt{-g}\left(\frac{R}{16\pi
G_N}+\frac1{g_o^2}\left(\frac12\phi
\Fc(\Box_g)\phi-V(\phi)\right)-\Lambda\right), \label{action_model2}
\end{equation}
where $G_N$ is the dimensionless gravitational constant: $ G_N=1/(8\pi M_P^2\alpha^{\prime})$, $M_P$
is the Planck mass and $\alpha^{\prime}$ is the string length squared.  The dimensionless parameter $g_o$
is the open string coupling constant divided on $\sqrt{\alpha^{\prime}}$. We use dimensionless
coordinates $x_\mu$ and the signature $(-,+,+,+)$.  $R$ is the scalar
curvature, $g_{\mu\nu}$ is the metric tensor, $\Lambda$ is a dimensionless constant.
 The potential is an
arbitrary quadratic polynomial: $V(\phi)=C_2\phi^2+C_1\phi+C_0$. The
Beltrami--Laplace (d'Alembert) operator $\Box_g$ is applied to scalar functions and
can be written as follows
\begin{equation}
\Box_g  = \frac{1}{\sqrt{-g}} \partial_\mu
\sqrt{-g}g^{\mu\nu}\partial_\nu \,.
\end{equation}

The function $\Fc$ is assumed to be an analytic function, therefore,
one can represent it by the convergent series expansion:
\begin{equation}
\Fc(\Box_g)=\sum\limits_{n=0}^{\infty}f_n\Box_g^{\;n},
\end{equation}
where $f_n$ are constants.
The function $\Fc$ may have infinitely many roots manifestly producing
thereby the nonlocality~\cite{noghosts,KV}. This model has been studied
in~\cite{Koshelev07,KV} with an additional condition that all roots of
the function $\Fc$ are simple. In this paper we consider double roots
as well. To clarify the interest to consider the case of double roots
let us study a trivial example with
\begin{equation}
\Fc(\Box_g)=(\Box_g-J_1)(\Box_g-J_2),
\end{equation}
where $J_1$ and $J_2$ are nonzero constants.

In the Minkowski space--time for $\phi$, depending only on time, we
obtain the following equation of motion
\begin{equation}
(\partial_t^2-J_1)(\partial_t^2-J_2)\phi(t)=0.
\end{equation}

This fourth order differential equation is equivalent to the following
system of two second order equations:
\begin{equation}
(\partial_t^2-J_1)\xi(t)=0,\qquad (\partial_t^2-J_2)\phi(t)=\xi(t).
\end{equation}

The first equation has the general solution
\begin{equation}
\xi(t)=B_1e^{\sqrt{J_1}t}+B_2e^{-\sqrt{J_1}t},
\end{equation}
where $B_1$ and $B_2$ are arbitrary constants. So, we get the following
second order equation in $\phi$
\begin{equation}
(\partial_t^2-J_2)\phi(t)=B_1e^{\sqrt{J_1}t}+B_2e^{-\sqrt{J_1}t}.
\end{equation}
In the non-resonance case (two simple roots $J_1$ and $J_2$) we get
\begin{equation}
\phi(t)=\frac{B_1}{J_1-J_2}e^{\sqrt{J_1}t}+\frac{B_2}{J_1-J_2}e^{-\sqrt{J_1}t}+
B_3e^{\sqrt{J_2}t}+B_4e^{-\sqrt{J_2}t},
\end{equation}
whereas in the resonance case (one double root $J_2=J_1$) the general
solution is
\begin{equation}
\phi(t)=\frac{B_1}{2\sqrt{J_1}}te^{\sqrt{J_1}t}-
\frac{B_2}{2\sqrt{J_1}}te^{-\sqrt{J_1}t}+B_3e^{\sqrt{J_1}t}+B_4e^{-\sqrt{J_1}t},
\end{equation}
where $B_3$ and $B_4$ are arbitrary constants.
 This trivial example shows that
behavior of solutions in the cases of one double root and two simple
roots are essentially different and one can not approximate double
roots by two simple roots, which are at a very small distance.
Resonance phenomenons are important and actively studied in various
domains of physics.

\section{Algorithm of
localization}

\subsection{Einstein equations}

From action (\ref{action_model2}) we obtain the following equations
\begin{eqnarray}
R_{\mu\nu}-\frac{1}{2}R g_{\mu\nu}&=&8\pi G_N\left(T_{\mu\nu}-\Lambda g_{\mu\nu}\right),
\label{EOJ_g}\\
\Fc(\Box_g)\phi&=&\frac{dV}{d\phi}, \label{EOJ_tau}
\end{eqnarray}
where $R_{\mu\nu}$ is the Ricci tensor.

The energy--momentum (stress) tensor $T_{\mu\nu}$, which is calculated
by the standard formula
\begin{equation}
T_{\mu\nu}={}-\frac{2}{\sqrt{-g}}\frac{\delta{S}}{\delta g^{\mu\nu}},
\end{equation}
can be presented in the following form:
\begin{equation}
\label{TEV}
T_{\mu\nu}=\frac{1}{g_o^2}\Bigl(E_{\mu\nu}+E_{\nu\mu}-g_{\mu\nu}\left(g^{\rho\sigma}
E_{\rho\sigma}+W\right)\Bigr),
\end{equation}
where
\begin{equation}
E_{\mu\nu}\equiv\frac{1}{2}\sum_{n=1}^\infty
f_n\sum_{l=0}^{n-1}\partial_\mu\Box_g^l\phi\partial_\nu\Box_g^{n-1-l}\phi,
\end{equation}
\begin{equation}
W\equiv\frac{1}{2}\sum_{n=2}^\infty
f_n\sum_{l=1}^{n-1}\Box_g^l\phi\Box_g^{n-l}\phi-\frac{f_0}{2}\phi^2+V(\phi).
\end{equation}

In the case of the zero potential $V(\phi)=0$, using the equation
\begin{equation}
 F(\Box_g)\phi=0,\quad \Longleftrightarrow \quad
f_0\phi={}-\sum\limits_{n=1}^{\infty}f_n\Box_g^{\;n}\phi,
\end{equation}
one can obtain that $W$ for $V(\phi)=0$ is equal to
\begin{equation}
W_0=\frac{1}{2}\sum_{n=1}^\infty
f_n\sum_{l=0}^{n-1}\Box_g^l\phi\Box_g^{n-l}\phi.
\end{equation}
The formula for energy--momentum tensor with $W_0$ has been proposed
in~\cite{Koshelev07} (see also~\cite{KV}).

The main idea of finding the solutions to the equations of motion is to
start with equation (\ref{EOJ_tau}) for $V(\phi)=0$ and to solve it,
assuming the function $\phi$ is an eigenfunction of the
Beltrami--Laplace operator $\Box_g$. If $\Box_g\phi=J\phi$, then such a
function $\phi$ is a solution to (\ref{EOJ_tau}) if and only if
\begin{equation}
\label{Betaequ}  \Fc(J)=0.
\end{equation}
The latter condition is known as the \textit{characteristic} equation.
Note that values of roots of $\Fc(J)$ do not depend on the metric. In
this paper we show how the case of an arbitrary quadratic potential
$V(\phi)$ can be analyzed with the help of roots of the function
$\Fc(J)$.

 Let
us denote simple roots of $\Fc$ as $J_i$ and double roots of $\Fc$ as
$\tilde{J}_k$.
 A particular solution of equation (\ref{EOJ_tau}) we  seek in the
 following form
 \begin{equation}
 \label{phi0}
 \phi_0=\sum\limits_{i=1}^{N_1}\phi_i+\sum\limits_{k=1}^{N_2}\tilde\phi_k,
\end{equation}
where
\begin{equation}
(\Box_g-J_i)\phi_i=0, \qquad (\Box_g-\tilde{J}_k)^2\tilde\phi_k=0.
\label{equphi}
\end{equation}

 The fourth order differential equation
$(\Box_g-\tilde{J_k})(\Box_g-\tilde{J_k})\tilde\phi_k=0$ is equivalent
to the following system of the second order equations:
\begin{equation}
(\Box_g-\tilde{J_k})\tilde\phi_k=\varphi_k,\qquad
(\Box_g-\tilde{J_k})\varphi_k=0.
\end{equation}

Without loss of generality we assume that for any $i_1$ and $i_2\neq
i_1$ conditions $J_{i_1}\neq J_{i_2}$ and
${\tilde{J}}_{i_1}\neq{\tilde{J}}_{i_2}$ are satisfied.

\subsection{Zero potential $V(\phi)$}

 It is convenient to consider the cases $C_1=0$ and
$C_1\neq 0$ separately. In this subsection we consider the case of zero
potential ($C_1=0$), the case of a linear potential is considered in
the next subsection.

Modifying values of $f_0$ and $\Lambda$, we can transform action
(\ref{action_model2}) with the potential $V(\phi)=C_2\phi^2+C_0$ to the
action with zero potential. So, without loss of generality, we can put
$C_2=0$ and $C_0=0$ and use the energy--momentum tensor for $\phi_0$,
which has been calculated in~\cite{Vernov2010}. It has been obtained
that for any analytical function $\Fc(J)$, which has simple roots $J_i$
and double roots $\tilde{J}_k$, and any $\phi_0$ given by (\ref{phi0})
the energy--momentum tensor
\begin{equation}
T_{\mu\nu}\left(\phi_0\right)=
T_{\mu\nu}\left(\sum\limits_{i=1}^{N_1}\phi_i+\sum\limits_{k=1}^{N_2}\tilde\phi_k\right)=
\sum\limits_{i=1}^{N_1}T_{\mu\nu}(\phi_i)+\sum\limits_{k=1}^{N_2}T_{\mu\nu}(\tilde\phi_k),
\label{Tmunugen}
\end{equation}
where all $T_{\mu\nu}$ are given by (\ref{TEV}) and
\begin{equation}
E_{\mu\nu}(\phi_i)=\frac{{
\Fc'(J_i)}}{2}\partial_{\mu}\phi_i\partial_{\nu}\phi_i,\qquad
W(\phi_i)=\frac{J_i \Fc'(J_i)}{2}\phi_i^2, \label{EWsimpleroot}
\end{equation}
\begin{equation}
\label{Edr} E_{\mu\nu}(\tilde\phi_k)= \frac{{
\Fc''(\tilde{J}_k)}}{4}\left(\partial_\mu\tilde\phi_k\partial_\nu\varphi_k+\partial_\nu\tilde\phi_k\partial_\mu\varphi_k\right)+
\frac{\Fc'''(\tilde{J}_k)}{12}\partial_\mu\varphi_k\partial_\nu\varphi_k,
\end{equation}
\begin{equation}
\label{Vdr} W(\tilde{\phi_k})=\frac{\tilde{J}_k
\Fc''(\tilde{J}_k)}{2}\tilde\phi_k\varphi_k+ \left(\frac{{\tilde{J}_k
\Fc'''(\tilde{J}_k)}}{12}+\frac{{
\Fc''(\tilde{J}_k)}}{4}\right)\varphi_k^2,
\end{equation}
where a prime denotes a derivative with respect to $J$: $\Fc'\equiv
\frac{d\Fc}{dJ}$, \ $\Fc''\equiv \frac{d^2\Fc}{dJ^2}$ and $\Fc'''\equiv
\frac{d^3 \Fc}{dJ^3}$. The result has been obtained for an arbitrary
metric.

Considering the following local action
\begin{equation}
S_{loc}=\int d^4x\sqrt{-g}\left(\frac{R}{16\pi
G_N}-\Lambda\right)+\sum_{i=1}^{N_1}S_i+\sum_{k=1}^{N_2}\tilde{S}_k,
\label{Sloc}
\end{equation}
where
\begin{equation}
S_i=\!{}-\frac{1}{g_o^2}\int d^4x\sqrt{-g}
\frac{\Fc'(J_i)}{2}\left(g^{\mu\nu}\partial_\mu\phi_i\partial_\nu\phi_i
+J_i\phi_i^2\right),
\end{equation}
\begin{equation}
\begin{array}{rl}
\!\displaystyle\tilde{S}_k=&\!\displaystyle\! {}-\frac{1}{g_o^2}\int
d^4x\sqrt{-g}\left(g^{\mu\nu}\left(\frac{{
\Fc''(\tilde{J}_k)}}{4}\left(\partial_\mu
\tilde{\phi}_k\partial_\nu\varphi_k+\partial_\nu
\tilde{\phi}_k\partial_\mu\varphi_k\right)+{}\right.\right.\\
\displaystyle +&\displaystyle \left.\frac{
\Fc'''(\tilde{J}_k)}{12}\partial_\mu\varphi_k\partial_\nu\varphi_k\right)+
\left. \frac{\tilde{J}_k \Fc''(\tilde{J}_k)}{2}\tilde\phi_k\varphi_k
+\left(\frac{{\tilde{J}_k \Fc'''(\tilde{J}_k)}}{12}+\frac{{
\Fc''(\tilde{J}_k)}}{4}\right)\varphi_k^2\right), \label{Slocdr}
\end{array}
\end{equation}
we can see that solutions of the Einstein equations and equations in
$\phi_k$, $\tilde{\phi}_k$ and $\varphi_k$, obtained from this action,
solves the initial system of nonlocal equations (\ref{EOJ_g}) and
(\ref{EOJ_tau}). Thus,  one can find special solutions of nonlocal
equations solving a system of local (differential)
equations.

To clarify physical interpretation of local fields $\tilde{\phi}_k$ and
$\varphi_k$, we diagonalize the kinetic terms of these scalar fields in
(\ref{Sloc}). Expressing $\tilde{\phi}_k$ and $\varphi_k$ in terms of
new fields $\xi_k$ and~$\chi_k$:
\begin{equation}
\begin{split}
    \tilde{\phi}_k&=\frac{1}{2 \Fc''(\tilde{J}_k)}\left(
    \left( \Fc''(\tilde{J}_k)-\frac{1}{3} \Fc'''(\tilde{J}_k)\right)
    \xi_k-\left( \Fc''(\tilde{J}_k)+\frac{1}{3} \Fc'''(\tilde{J}_k)\right)\chi_k\right),\\
    \varphi_k&=\xi_k+\chi_k,
\end{split}
\end{equation}
we obtain the corresponding $\tilde{S}_k$ in the following form:
\begin{equation}
\begin{array}{l}
\displaystyle\tilde{S}_k={}-\frac{1}{g_o^2}\int
d^4x\sqrt{-g}\left(g^{\mu\nu}\frac{ \Fc''(\tilde{J}_k)}{4}\left(\partial_\mu
\xi_k\partial_\nu\xi_k-\partial_\nu
\chi_k\partial_\mu\chi_k\right)+{}\right.\\
\displaystyle {}+\frac{\tilde{J}_k}{4}\left(\left(
\Fc''(\tilde{J}_k)-\frac{1}{3} \Fc'''(\tilde{J}_k)\right)
    \xi_k-\left( \Fc''(\tilde{J}_k)+\frac{1}{3} \Fc'''(\tilde{J}_k)\right)\chi_k\right)(\xi_k+\chi_k)+{}\\
\displaystyle {}+ \left. \left(\frac{{\tilde{J}_k
\Fc'''(\tilde{J}_k)}}{12}+\frac{{
\Fc''(\tilde{J}_k)}}{4}\right)(\xi_k+\chi_k)^2\right).
\end{array}
\end{equation}

It is easy to see that each $\tilde{S}_k$ includes one phantom scalar
field and one standard scalar field. So, in the case of one double root
we obtain a quintom model. In the Minkowski space appearance of phantom
fields in models, when $ \Fc(\Box)$ has a double root, has been
obtained in~\cite{PaisU}.  If $\Fc(J)$ has simple real roots, then
positive and negative values of $\Fc'(J_i)$ alternate, so we can obtain
phantom fields and, in the case of two simple roots, a quintom model.

\textbf{Remark 1.} If $ \Fc(J)$ has an infinity number of roots then
one nonlocal model corresponds to infinity number of different local
models. In this case the initial nonlocal action (\ref{action_model2})
generates infinity number of local actions (\ref{Sloc}).

\textbf{Remark 2.} We should prove that the way of localization is
self-consistent. To construct local action (\ref{Sloc}) we assume that
equations (\ref{equphi}) are satisfied. Therefore, the method of
localization is correct only if these equations can be obtained from
the local action $S_{loc}$. The straightforward calculations show that
\begin{equation}
\frac{\delta{S_{loc}}}{\delta \phi_i}=0 \quad \Leftrightarrow \quad
\Box_g\phi_i=J_i\phi_i; \qquad \frac{\delta{S_{loc}}}{\delta
\tilde{\phi}_k}=0 \quad \Leftrightarrow \quad
\Box_g\varphi_k=\tilde{J}_k\varphi_k. \label{equvarphi}
\end{equation}
Using (\ref{equvarphi}) we obtain
\begin{equation}
\frac{\delta{S_{loc}}}{\delta \varphi_k}=0 \qquad \Leftrightarrow
\qquad \Box_g\tilde{\phi}_k=\tilde{J}_k\tilde{\phi}_k+\varphi_k.
\end{equation}

So, the way of localization is self-consistent in the case of $ \Fc(J)$
with simple and double roots~\cite{Vernov2010}. The self-consistence of
similar approach for $ \Fc(J)$ with only simple roots has been proven
in~\cite{AJV0711,KV}.

In spite of the above-mention equations we obtain  from $S_{loc}$ the
Einstein equations:
\begin{equation}
G_{\mu\nu}=8\pi G_N\left(T_{\mu\nu}(\phi_0)-\Lambda g_{\mu\nu}\right),
\end{equation}
where $\phi_0$ is given by (\ref{phi0}) and $T_{\mu\nu}(\phi_0)$ can be
calculated by (\ref{Tmunugen}).

So, we have obtained such systems of differential equations that any
solutions of these systems are particular solutions of the initial
nonlocal equations (\ref{EOJ_g}) and (\ref{EOJ_tau}).

\subsection{Linear potential $V(\phi)$}

Let us consider the model with action (\ref{action_model2}) in the case
$C_1\neq 0$. For the string field theory inspired form of $\Fc(\Box)$
the case $f_0\neq 0$ has been considered in~\cite{MN}. In this case the
effective potential: $-f_0\phi^2/2+V(\phi)+\Lambda$, is a quadratic
potential. Using the condition $f_0\neq 0$, we boil down the case with
an arbitrary $C_1$ to the case with $C_1=0$. Indeed, we work in a new
scalar field
\begin{equation}
\label{chi} \chi=\phi-\frac{C_1}{f_0}
\end{equation}
and get the energy--momentum tensor in the form (\ref{TEV}) with
\begin{equation}
E_{\mu\nu}=\frac{1}{2}\sum_{n=1}^\infty
f_n\sum_{l=0}^{n-1}\partial_\mu\Box_g^l\chi\partial_\nu\Box_g^{n-1-l}\chi,
\end{equation}
\begin{equation}
W=\frac{1}{2}\sum_{n=1}^\infty
f_n\sum_{l=1}^{n-1}\Box_g^l\chi\Box_g^{n-l}\chi-\frac{f_0}{2}\chi^2+\frac{C_1^2}{2f_0}.
\end{equation}

It is easy to see that
\begin{equation}
\label{linearEq} \Fc(\Box) \phi = C_1 \qquad \Longleftrightarrow \qquad
\Fc(\Box)\chi=0.
\end{equation}
The constant $C_1^2/(2f_0)$ can be consider as a part of the
cosmological constant. Thus, in terms of $\chi$ we obtain a model
without linear term and can conclude that at $f_0\neq 0$ the adding of
a linear term to the potential shifts the scalar field on the constant
and changes the value of the cosmological constant.

Let us consider the case $f_0=0$. In this case   $J=0$ is a root of the
characteristic equation (\ref{Betaequ}). It is easy to show, that the
function
\begin{equation}
 \tilde{\chi}=\phi_0+\psi,
\end{equation}
where $\phi_0$ and $\psi$ are solutions of the following equations
\begin{equation}
\Fc(\Box) \phi_0 =0, \qquad \Box^m\psi=\frac{C_1}{f_m},
\end{equation}
$m$ is the order of the root $J=0$, satisfies
\begin{equation}
\Fc(\Box) \tilde{\chi} = C_1.
\end{equation}
The function $\phi_0$ is given by (\ref{phi0}), but  the sum do not
include $\phi_{i_0}$, which corresponds to the root $J=0$, because this
function can not be separated from $\psi$. We consider the cases of
$m=1$ and $m=2$. In the last case, when $J=0$ is a double root, we
denote the function $\psi$ as $\tilde{\psi}$.

To localize the Einstein equations one should calculate the
energy--momentum tensor for~$\tilde{\chi}$:
\begin{equation}
T_{\mu\nu}(\tilde{\chi})=T_{\mu\nu}(\psi)+T_{\mu\nu}(\phi_0)+T_{\mu\nu}^{cr}(\psi,\phi_0).
\end{equation}

Let us calculate
\begin{equation}
W(\tilde{\chi})=\frac{1}{2}\sum_{n=2}^\infty
f_n\sum_{l=1}^{n-1}\Box_g^l\tilde{\chi}\Box_g^{n-l}\tilde{\chi}+C_1\tilde{\chi}.
\end{equation}

To simplify notation we choose $\phi_0=\phi_i$, where $J_i$ is a simple
root, the generalization to an arbitrary $\phi_0$ is straightforward.
In the case of the simple root $J=0$ we have $\Box\psi=C_1/f_1$ and
\begin{equation}
W(\psi+\phi_i)=\frac{f_2C_1^2}{2f_1^2}+W(\phi_i)+\sum_{n=2}^\infty
\frac{C_1}{f_1}f_nJ_i^{n-1}\phi_i+C_1(\psi+\phi_i).
\end{equation}
Using
\begin{equation}
\sum_{n=2}^\infty
\frac{C_1}{f_1}f_nJ_i^{n-1}\phi_i=\frac{C_1}{f_1J_i}\sum_{n=1}^\infty
f_nJ_i^{n}\phi_i-C_1\phi_i=\left(\frac{C_1\Fc(J_i)}{f_1J_i}-C_1\right)\phi_i=-C_1\phi_i,
\end{equation}
we obtain
\begin{equation}
W(\psi+\phi_i)=W(\psi)+W(\phi_i),
\end{equation}
where $W(\phi_i)$ is given by (\ref{EWsimpleroot}) and
\begin{equation}
W(\psi)=C_1\psi+\frac{f_2C_1^2}{2f_1^2}.
\end{equation}
Similar calculations give
\begin{equation}
E_{\mu\nu}(\tilde{\chi})=E_{\mu\nu}(\psi)+E_{\mu\nu}(\phi_0),
\end{equation}
where
\begin{equation}
E_{\mu\nu}(\psi)=\frac12
f_1\partial_\mu\psi\partial_\nu\psi.
\end{equation}

The function $\phi_0$ is given by  (\ref{phi0}) and satisfies equation
(\ref{EOJ_tau}) with $C_1=0$, therefore, we use $W_0$ instead of $W$ to
calculate $T_{\mu\nu}(\phi_0)$ and obtain equality (\ref{Tmunugen}).

So, we get
\begin{equation}
T_{\mu\nu}^{cr}(\psi,\phi_0)=0 \qquad \mbox{and} \qquad
T_{\mu\nu}(\tilde{\chi})=T_{\mu\nu}(\psi)+T_{\mu\nu}(\phi_0).
\end{equation}

In the case of the double root $J=0$ equation in $\tilde{\psi}$ is as follows
\begin{equation}
\Box^2\tilde{\psi}=\frac{C_1}{f_2}, \ \Longleftrightarrow \
\left\{
\begin{array}{l}
\displaystyle \Box\tilde{\psi}=\tau, \\[2.7mm]
\displaystyle \Box\tau=\frac{C_1}{f_2}.
\end{array}
\right.
 \end{equation}
We obtain
\begin{equation}
T_{\mu\nu}(\tilde{\chi})=T_{\mu\nu}(\tilde{\psi})+T_{\mu\nu}(\phi_0),
\end{equation}
\begin{equation}
E_{\mu\nu}(\tilde{\psi})=\frac{1}{2}\left(f_2
\left(\partial_\mu\tilde{\psi}\partial_\nu\tau+\partial_\nu\tilde{\psi}\partial_\mu\tau\right)+
f_3\partial_\mu\tau\partial_\nu\tau\right),
\end{equation}
\begin{equation}
W(\tilde{\psi})=\frac{f_2}{2}\tau^2+C_1\tilde{\psi}+\frac{f_3C_1}{f_2}\tau.
\end{equation}

The obtained formulae allow to generalize the algorithm of
localization, proposed in~\cite{Vernov2010} to the case $C_1\neq 0$.
 For an arbitrary
quadratic potential $V(\phi)=C_2\phi^2+C_1\phi+C_0$ there exists the
following algorithm of localization:
\begin{itemize}

\item Change values of $f_0$ and $\Lambda$ such that the potential
takes the form $V(\phi)=C_1\phi$.

\item Find roots of the function $ \Fc(J)$ and calculate orders of
them.

\item Select a finite number of simple and double roots.

\item Construct the corresponding local action. In the case $C_1=0$ one
should use formula (\ref{Sloc}). In the case $C_1\neq 0$ and $f_0\neq
0$ one should use (\ref{Sloc}) with the replacement of the scalar field
$\phi$ by $\chi$ (formula (\ref{chi})) and the corresponding
modification of the cosmological constant. In the case $C_1\neq 0$ and
$f_0=0$ the local action is the sum of (\ref{Sloc}) and either
\begin{equation*}
S_{\psi}=\!{}-\frac{1}{2g_o^2}\int\! d^4x\sqrt{-g}\left[
f_1g^{\mu\nu}\partial_\mu\psi\partial_\nu\psi+2C_1\psi+\frac{f_2C_1^2}{f_1^2}\right],
\end{equation*}
in the case of simple root $J=0$, or
\begin{equation*}
\! S_{\tilde{\psi}}=\frac{-1}{2g_o^2}\!\int\!\! d^4x\sqrt{-g}\left[
g^{\mu\nu}\!\left(f_2(\partial_\mu\tilde{\psi}\partial_\nu\tau
+\partial_\nu\tilde{\psi}\partial_\mu\tau)+f_3\partial_\mu\tau\partial_\nu\tau\right)\!
+f_2\tau^2+2C_1\tilde{\psi}+\frac{f_3C_1}{2f_2}\tau\!\right]\!,\!
\end{equation*}
in the case of double root $J=0$. Note that in the case $C_1\neq 0$ and
$f_0=0$ the local action (\ref{Sloc}) has no term, which corresponds to
the root $J=0$.

\item Vary the obtained local action and get a system of the Einstein
equations and equations of motion. The obtained system is a finite
order system of differential equations, \textit{i.e.} we get a local
system.

\item Seek solutions of the obtained local system.

\end{itemize}

\section{Exact solutions}

\subsection{Root of $\Fc(\Box)$ in the case of the SFT inspired models}

The particular forms of  $\Fc(\Box_g)$ are inspired by the fermionic
SFT and the most well understood process of tachyon condensation.
Namely, starting with a non-supersymmetric configuration the tachyon of
the fermionic string rolls down towards the nonperturbative minimum of
the tachyon potential. This process represents the non-BPS brane decay
according to Sen's conjecture (see \cite{review-sft} for details). From
the point of view of the SFT the whole picture is not yet known and
only vacuum solutions were constructed. An effective field theory
description explaining the rolling tachyon in contrary is known and
numeric solutions describing the tachyon dynamics were
obtained~\cite{AJK}. This effective field theory description does
capture the nonlocality of the SFT. Linearizing the latter lagrangian
around the true vacuum one gets a model which is of main concern in the
present paper. The SFT inspired form of the function $\Fc(\Box_g)$,
which has the nonlocal operator $\exp(\Box_g)$ as a key ingredient:
\begin{equation}
\label{F} \Fc_{SFT}(\Box_g)=\xi^2 \Box_g+1-c\:e^{2\Box_g},
\end{equation}
where $\xi$ is a real parameter and $c$ is a positive constant, has
been considered in~\cite{AJV0701,AJV0711,MN}. The form of the term
$(e^{\square_g}\phi)^2$ is analogous to the form of the interaction
term for the tachyon field in the SFT action.

The characteristic equation $\Fc_{SFT}(J)=0$ has the following
solutions:
\begin{equation}
J_n=
-\frac{1}{2\xi^2}\left(2+\xi^2W_n\left(-\frac{2c}{\xi^2}e^{-2/\xi^2}\right)\right),
\end{equation}
where $n$ is an integer number, $W_n$ is the $n$-s branch of the
Lambert function satisfying a relation $W(z)e^{W(z)}=z$. The Lambert
function is a multivalued function, so $\Fc_{SFT}(J)$ has an infinite
number of roots. Parameters $\xi$ and $c$ are real, therefore if $J_n$
is a root of $\Fc_{SFT}(J)$, then the complex adjoined number $J_n^*$
is a root as well.

 If $J=\tilde{J}_0$ is a multiple root, then at this point
$\Fc_{SFT}(J)=0$ and $\Fc_{SFT}'(J)=0$. These equations give that
\begin{equation}
\label{z-0} \tilde{J}_0=\frac12-\frac{1}{\xi^2},
\end{equation}
 hence the root $\tilde{J}_0$ is a real number.
$\tilde{J}_0$ is a double roots because:
\begin{equation}
\Fc_{SFT}''(\tilde{J}_0)={}-4ce^{2\tilde{J}_0}\neq 0.
\end{equation}
The function $\Fc_{SFT}(J)$ has a double root if and only if
$c=\frac{\xi^2}{2}\:e^{(2/\xi^2-1)}$.

Roots of $\Fc_{SFT}(J)$ do not depend on metric. In the Minkowski
space-time these roots have been studied in~\cite{AJV0701}. The
function $\Fc_{SFT}$ always has an infinity number of complex roots.
Let us consider real roots of $\Fc_{SFT}$. There are three different
cases:

\begin{itemize}
\item If $c<1$, then for any values $\xi$ the function $\Fc_{SFT}(J)$
has two simple real root: one is positive, another is negative.

\item If $c=1$, then $J=0$ is a simple root at $\xi^2\neq 2$. A
positive root exist if and only if $\xi^2>2$. At $\xi^2<2$ a  negative
root exist. If $\xi^2=2$, then $J=0$ is a double root.

\item If $c>1$, then $\Fc_{SFT}(J)$ has
\begin{itemize}
\item two negative simple roots for $\xi^2<\xi^2_1$,

\item a negative double root for $\xi^2=\xi^2_1$,

\item no real roots for $\xi^2_2>\xi^2>\xi^2_1$,

\item a positive double root for $\xi^2=\xi^2_2$,

\item two real positive roots for $\xi^2>\xi^2_2$, where
\begin{equation}
\label{xi-max} \xi_1^2={}-\frac{2}{W_{-1}({}-exp(-1)/c)} ,\qquad
\xi_2^2={}-\frac{2}{W_{0}({}-exp(-1)/c)}.
\end{equation}
\end{itemize}
\end{itemize}

\begin{figure}[h]
\centering
\includegraphics[width=47.2mm]{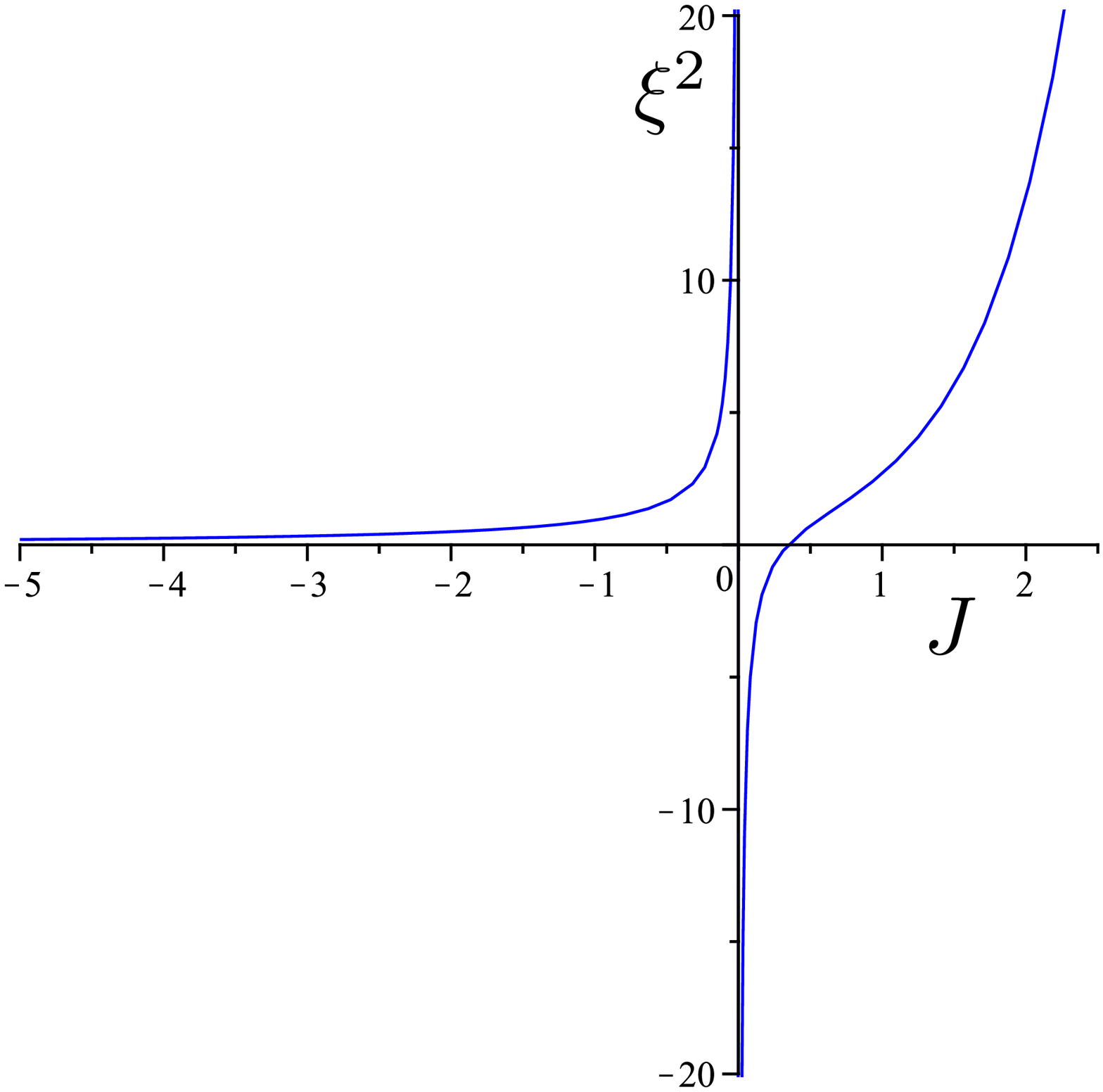} \ \ \ \ \
\includegraphics[width=47.2mm]{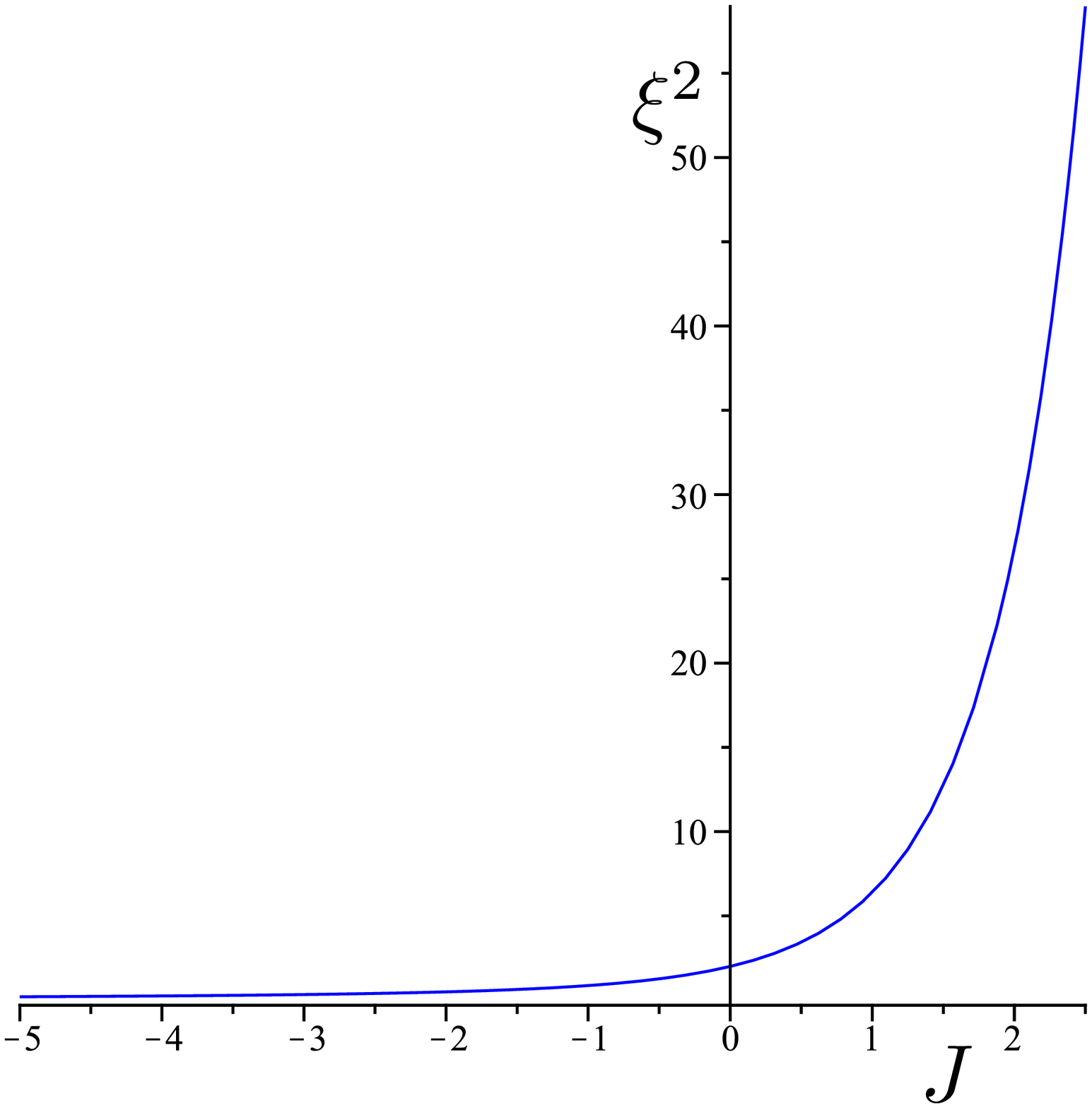} \ \ \ \ \
\includegraphics[width=47.2mm]{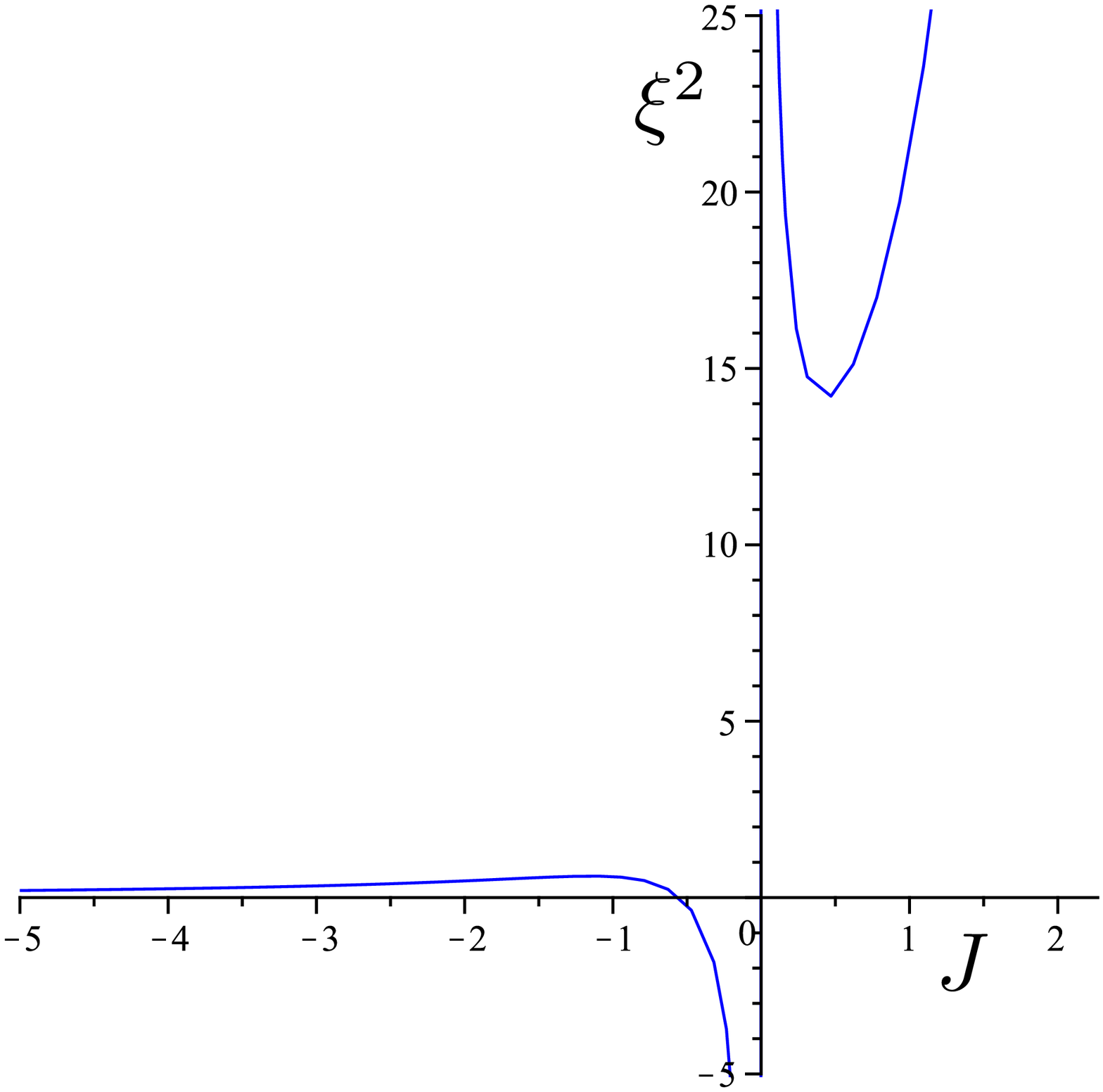}
\caption{The dependence of the function $\tilde{g}(J,c)$, which is equal to
$\xi^2$, on $J$ at $c=1/2$ (left), $c=1$ (center) and $c=3$ (right).}
\label{xi2_m}
\end{figure}

To illustrate the dependence of the parameter $\xi^2$ on real roots we
introduce the function $\tilde{g}(J,c)$:
\begin{equation}\label{equxi}
     \tilde{g}(J,c)=\xi^2=\frac{c\,e^{2J}-1}{J},
\end{equation}
and plot $\tilde{g}(J,c)$ as a function of $J$ at three different values of $c$
(see Figure \ref{xi2_m}).

Let us consider a special values of $\xi^2$ and $c$, which have been
obtained in the SFT inspired cosmological model. From the action for
the tachyon in the SFT~\cite{TCSFT} the following equation has been
obtained~\cite{AK}:
\begin{equation}
 (-\xi_0^2\tilde{\alpha}^2+1)=3e^{-\tilde{\alpha}^2/4},
\end{equation}
where
\begin{equation}
\xi_0^2={}-\frac{1}{4\ln\left(\frac{4}{3\sqrt{3}}\right)}\approx
0.9556.
\end{equation}

Substituting $J=-\tilde{\alpha}^2/8$, we obtain $\Fc_{SFT}$  with
$\xi_{SFT}^2=8\xi_0^2$ and $c=3$. At $c=3$ we obtain that $\xi^2_1=
0.6080355395$ and $\xi^2_2= 14.16157383$. Therefore,
$\xi^2_2>\xi_{SFT}^2>\xi^2_1$, so there exists no real root at these
values of parameters.

\subsection{Exact Solution in the Friedmann--Robertson--Walker metric}

Let us consider the Einstein equations, which corresponds to a real
simple root $J_1$ in the Friedmann--Robertson--Walker
metric~\cite{AJV0711}:
\begin{equation}
\left\{
\begin{array}{l}
\displaystyle 3H^2= \frac{4\pi G_N {\cal
F}'(J_1)}{g_o^2}\left(\dot\phi^2+
    J_1\phi^2\right)+8\pi G_N\Lambda,
\\[7.2mm]
\displaystyle \dot H={}-\frac{4\pi G_N {\cal F}'(J_1)}{g_o^2}\dot\phi^2,
\end{array}
\right. \label{eomprholocal1}
\end{equation}
where a dot denotes a time derivative.

Exact real solutions of this system have been obtained
in~\cite{AVzeta,AJV0711}. In our notations these solutions are as
follows:

At $J_1 > 0$
\begin{equation}
  \phi(t)={}\pm \frac{\sqrt{3J_1}
g_o^2}{6\pi G_N {\cal F}'(J_1)}(t-t_0),\qquad  H(t)={}-\frac{J_1
g_o^2}{6\pi G_N {\cal F}'(J_1)}(t-t_0),
\end{equation}
where $t_0$ is an arbitrary constant. These solutions exist only at
\begin{equation}
\Lambda ={} -\frac{J_1g_o^2}{24\pi^2  G_N^2 {\cal F}'(J_1)}\, .
\end{equation}

 At $J_1=0$ summing the first and the second equations of
(\ref{eomprholocal1}), we obtain:
\begin{equation}
\dot H=8\pi G_N\Lambda-3H^2.
\end{equation}

The type of solution depends on sign of $\Lambda$:

\begin{itemize}
\item $\Lambda=0$
\begin{equation}
H(t)={}-\frac{1}{3(t-t_0)},    \quad  \phi(t)={\tilde C}_1\pm
\frac{\sqrt{3} g_o}{\sqrt{\pi G_N {\cal F}'(0)}}\ln(t-t_0),
\end{equation}
where $t_0$ and ${\tilde C}_1$ are arbitrary constants.

\item If $\Lambda>0$, then we obtain solutions:
\begin{equation}
\label{Hnew} H_1(t)=\frac{2\sqrt{6\pi
G_N\Lambda}}{3}\tanh\left(2\sqrt{6\pi G_N\Lambda}(t-t_0)\right),
\end{equation}
\begin{equation}
\phi_1(t)=\pm\sqrt{\frac{{}-g_o^2}{12\pi G_N {\cal F}'(0)}}\arctan
\left(\sinh\left(2\sqrt{6\pi G_N \Lambda}(t-t_0)\right)\right)+{\tilde
C}_2
\end{equation}
and
\begin{equation}
\label{Hnew2} \tilde{H}_1(t)=\frac{2\sqrt{6\pi
G_N\Lambda}}{3}\coth\left(2\sqrt{6\pi G_N\Lambda}(t-t_0)\right),
\end{equation}
\begin{equation}
\tilde{\phi}_1(t)=\pm\sqrt{\frac{g_o^2}{12 \pi G_N {\cal
F}'(0)}}\ln\left(\tanh\left(\sqrt{6\pi
G_N\Lambda}\left(t-t_0\right)\right)\right)+{\tilde C}_2,
\end{equation}

hereafter $t_0$ and ${\tilde C}_2$ are arbitrary real constants.

\item In the case $\Lambda<0$ we obtain the following real solution:
\begin{equation}
H_2(t)={}-\frac{2\sqrt{-6\pi G_N\Lambda}}{3}\tan\left(2\sqrt{-6\pi
G_N\Lambda}(t-t_0)\right),
\end{equation}
\begin{equation}
\phi_2(t)={}\pm \sqrt{\frac{g_o^2}{12\pi G_N {\cal F}'(0)}}
\mathrm{arctanh}\left(\sin\left(2\sqrt{-6\pi
G_N\Lambda}(t-t_0)\right)\right)+{\tilde C}_2.
\end{equation}

\end{itemize}

The stability of the exact solutions, obtained in the
Friedmann--Robertson--Walker metric~\cite{AJV0711}, has been analysed
in~\cite{ABJV0903}.

\subsection{Exact solutions in the  Bianchi I metric}

In Bianchi I metric with the interval
\begin{equation}
{ds}^{2}={}-{dt}^2+a_1^2(t)dx_1^2+a_2^2(t)dx_2^2+a_3^2(t)dx_3^2,
\end{equation}
the Einstein equations, which correspond to the simple root $J=0$ have
the following form:
\begin{equation}
\label{a} H_1H_2+H_1H_3+H_2H_3=8\pi
G_N\left(\frac{\Fc'(0)}{2g_o^2}\dot{\phi}^2+\Lambda\right),
\end{equation}
\begin{equation}
\label{b} \dot H_2+H_2^2+\dot H_3+H_3^2+H_2H_3={}-8\pi
G_N\left(\frac{\Fc'(0)}{2g_o^2}\dot{\phi}^2-\Lambda\right),
\end{equation}
\begin{equation}
\label{c} \dot H_1+H_1^2+\dot H_2+H_2^2+H_1H_2={}-8\pi
G_N\left(\frac{\Fc'(0)}{2g_o^2}\dot{\phi}^2-\Lambda\right),
\end{equation}
\begin{equation}
\label{d} \dot H_1+H_1^2+\dot H_3+H_3^2+H_1H_3={}-8\pi
G_N\left(\frac{\Fc'(0)}{2g_o^2}\dot{\phi}^2-\Lambda\right),
\end{equation}
where $H_k\equiv \dot a_k/a_k$, $k=1,2,3$. Note that $\Fc'(0)\neq 0$.

Our goal is to find exact solutions to system (\ref{a})--(\ref{d}). Of
course, there exist isotropic solutions, which coincide with exact
solutions in the Friedmann--Robertson--Walker metric. For those
solutions $H_1(t)=H_2(t)=H_3(t)$. At the same time exact anisotropic
solutions do exist.

For $\Lambda=0$ we obtain the following solution:
\begin{equation}
\label{BianchiH_Lambda0}
H_1(t)=\frac{\tilde{C}_2+\tilde{C}_1+1}{\tilde{C}_2 (t-t_0)},\quad
H_2(t)={}-\frac{\tilde{C}_1}{\tilde{C}_2 (t-t_0)},\quad
H_3(t)={}-\frac{1}{\tilde{C}_2 (t-t_0)}\, ,
\end{equation}
\begin{equation}
    \phi(t)={}\pm\frac{\sqrt{-\pi G_N \Fc'(0)\left({\tilde C}_1{\tilde C}_2
    +{\tilde C}_1^2+{\tilde C}_1+{\tilde C}_2+1\right)}}{4\pi G_N g_o \Fc'(0)
    {\tilde C}_2}\ln\left({\tilde C}_2(t-t_0)^2\right)+{\tilde C}_3,
\end{equation}
where  ${\tilde C}_1$, ${\tilde C}_2$, ${\tilde C}_3$, and  $t_0$ are
arbitrary constants.

For  $\Fc'(0)<0$ we obtain that $\phi(t)$ is a real function at
\begin{equation}
   {\tilde C}_1 \geqslant -1,\quad {\tilde C}_2>0 \quad\mbox{or}\quad {\tilde C}_1<-1,\quad
   {}-\frac{{\tilde C}_1^2+{\tilde C}_1+1}{{\tilde C}_1+1}>{\tilde C}_2>0.
\end{equation}

For $\Fc'(0)>0$ we obtain that $\phi(t)$ is a real function at
\begin{equation*}
{\tilde C}_1<-1,\qquad {\tilde C}_2>{}-\frac{{\tilde C}_1^2+{\tilde
C}_1+1}{{\tilde C}_1+1}.
\end{equation*}

Let us consider the case of $\Lambda=1/(8\pi G_N)$.
There exist not only the following isotropic solution
\begin{equation}
    H_1(t)=H_2(t)=H_3(t)=\frac{1}{\sqrt{3}}\tanh\left(\sqrt{3}(t-t_0)\right),
\end{equation}
but also an anisotropic one

\begin{equation}
\label{Anisol}
\begin{split}
    H_1(t)&=\frac{1}{\sqrt{3}}\tanh\left(\frac{\sqrt{3}}{2}(t-t_0)\right),\\
    H_2(t)&=\frac{1}{\sqrt{3}}\coth\left(\frac{\sqrt{3}}{2}(t-t_0)\right),\\
    H_3(t)&=\frac{1}{2\sqrt{3}}\left(\tanh\left(\frac{\sqrt{3}}{2}(t-t_0)\right)
    +\coth\left(\frac{\sqrt{3}}{2}(t-t_0)\right)\right).
\end{split}
\end{equation}
The corresponding scalar field is real at $\Fc'(0)>0$ and is equal to
\begin{equation*}
    \tilde{\phi}(t)={\tilde C}_4\pm\frac{1}{3\sqrt{2\pi G_N g_o^2 \Fc'(0)}}
    \left(\ln\left(e^{\sqrt{3}(t-t_0)}+1\right)-\ln\left(e^{\frac{\sqrt{3}}{2}(t-t_0)}-1\right)
    -\ln\left(e^{\frac{\sqrt{3}}{2}(t-t_0)}+1\right)\right),
\end{equation*}
where ${\tilde C}_4$ is an arbitrary real constant.

\section{Conclusion}

The main result of this paper is the generalization of the algorithm of
localization on nonlocal models with linear potentials. This algorithm
is proposed for an arbitrary analytic function $\Fc(\Box_g)$, which has
both simple and double roots. We have proved that the same functions
solve the initial nonlocal Einstein equations and the obtained local
Einstein equations. We have found the corresponding local actions and
proved the self-consistence of our approach.

It is interesting to consider nonlocal models with an arbitrary
analytic $\Fc(\Box_g)$, without any restrictions on order of roots. The
consideration of simple and double roots allows us to make the
conjecture that the existence of local actions, which correspond to a
nonlocal action, does not depend on order of $\Fc(\Box_g)$ roots and
the method of finding particular solutions of the nonlocal Einstein
equations can be generalized on a nonlocal action with an arbitrary
analytic~$\Fc(\Box_g)$.

In the case of simple roots exact solutions in the
Friedmann--Robertson--Walker metric have been found in~\cite{AJV0711}
(their stability is considered in~\cite{ABJV0903}). In this paper we
present exact solutions in Friedmann--Robertson--Walker and Bianchi I
metrics. The algorithm of localization does not depend on metric, so it
can be used  to find solutions in other metrics. For example, the
well-known Fisher solutions~\cite{Fisher} (see
also~\cite{AboutFisher1,AboutFisher2}), which are static spherically
symmetric solutions for gravitational system with a massless scalar
field, are solutions of the nonlocal Einstein equations
(\ref{EOJ_g})--(\ref{EOJ_tau}) for any $\Fc(J)$, which has a simple
root $J_0=0$.


\section*{Acknowledgements}

The author is grateful to the organizers of the Dubna International
SQS'09 Workshop ("Supersymmetries and Quantum Symmetries--2009", Dubna,
Russia, July 29 --- August 3, 2009) for hospitality and financial
support. The author is grateful to I.Ya.~Aref'eva, A.S.~Koshelev,
N.~Nunes and A.F.~Zakharov for useful and stimulating discussions. This
research is supported in part by RFBR grant 08-01-00798, grants of
Russian Ministry of Education and Science NSh-1456.2008.2 and NSh-4142.2010.2 and by
Federal Agency for Science and Innovation under state contract
02.740.11.0244.


\begin{thebibliography}{72}
{\small
\bibitem{review-sft}  K.~Ohmori,\textit{ A Review on Tachyon Condensation in Open String Field Theories},
{\tt hep-th/0102085};\\
 I.Ya.~Aref'eva, D.M.~Belov, A.A.~Giryavets, A.S.~Koshelev and
P.B.~Medvedev, \textit{Noncommutative Field Theories and (Super)String Field Theories}, {\tt hep-th/0111208};\\
 W. Taylor, \textit{Lectures on D-branes, tachyon condensation, and string field theory},
 {\tt hep-th/0301094}
\bibitem{padic}
L.~Brekke, P.G.O.~Freund, M.~Olson, and E.~Witten,
\textit{Nonarchimedean
String Dynamics}, Nucl. Phys. \textbf{B 302} (1988) 365--402;\\
P.H. Frampton, Ya. Okada, \textit{Effective Scalar Field Theory of
$P$-Adic String},
Phys. Rev. \textbf{D 37} (1988) 3077--3079;\\
V.S.~Vladimirov, I.V.~Volovich, E.I.~Zelenov, \textit{$p$-adic Analysis
and Mathe\-matical Physics}, WSP, Singapore,
1994;\\
B. Dragovich, A.Yu. Khrennikov, S.V. Kozyrev, I.V. Volovich,
\textit{$p$-Adic Mathematical Physics}, Anal. Appl. \textbf{1} (2009)
1--17, {\tt arXiv:0904.4205}
\bibitem{IA1} I.Ya. Aref'eva, \textit{Nonlocal String Tachyon as
a Model for Cosmological Dark Energy}, AIP Conf. Proc. \textbf{826},
\textit{p-Adic Mathematical Physics}, eds. A.Yu. Khrennikov,
Z. Raki´c, I.V. Volovich, AIP, Melville, NY, 2006, pp.~301--311; {\tt astro-ph/0410443};\\
I.Ya.~Aref'eva, {\it D-brane as a Model for Cosmological Dark Energy},
in: "Contents and Structures of the Universe",
            eds. C. Magneville, R. Ansari, J. Dumarchez,
             and J.T.T. Van, \textit{Proc. of the XLIst Rencontres de
Moriond}, 2006, pp.~131--135;\\
I.Ya.~Aref'eva, \textit{Stringy Model of Cosmological Dark Energy}, AIP
Conf. Proc. \textbf{957}, \textit{Particles, Strings, and Cosmology},
eds. A. Rajantie, P. Dauncey, C. Contaldi, H. Stoica, AIP, Melville,
NY, 2007, pp.~297--300, {\tt arXiv:0710.3017}
\bibitem{AJ} I.Ya. Aref'eva and  L.V. Joukovskaya,
\textit{Time Lumps in Nonlocal Stringy Models and Cosmological
Applications}, JHEP \textbf{0510} (2005) 087, {\tt hep-th/0504200}
\bibitem{Calcagni}  G. Calcagni, \textit{Cosmological tachyon from cubic string field theory},
 JHEP \textbf{0605} (2006) 012,  {\tt hep-th/0512259}
\bibitem{Barnaby} N. Barnaby, T. Biswas, and  J.M. Cline,
\textit{p-adic Inflation},  JHEP \textbf{0704} (2007) 056,
{\tt hep-th/0612230}\\
 N. Barnaby and  J.M. Cline, \textit{Large Nongaussianity from Nonlocal
Inflation}, JCAP \textbf{0707} (2007) 017,
{\tt arXiv:0704.3426}\\
 N. Barnaby, \textit{Nonlocal Inflation}, Can. J. Phys. \textbf{87} (2009) 189--194,
{\tt arXiv:0811.0814}
\bibitem{Koshelev07} A.S. Koshelev, \textit{Non-local SFT Tachyon and Cosmology},
JHEP \textbf{0704} (2007) 029, {\tt hep-th/0701103}
\bibitem{AJV0701} I.Ya. Aref'eva, L.V. Joukovskaya, and S.Yu. Vernov,
 \textit{Bouncing and accelerating solutions in nonlocal stringy models}
 JHEP \textbf{0707} (2007) 087, {\tt hep-th/0701184}
\bibitem{AVzeta}
 I.Ya. Aref'eva and  I.V.  Volovich,  \textit{Quantization of the Riemann
 Zeta-Function and Cosmology},  Int. J. of Geom. Meth.
Mod. Phys.  \textbf{4} (2007) 881--895, {\tt hep-th/0701284}
\bibitem{Lidsey07} J.E. Lidsey,  \textit{Stretching the Inflaton
Potential with Kinetic Energy},  Phys. Rev.  {\bf D 76} (2007) 043511,
{\tt hep-th/0703007}
\bibitem{Calcagni07}
 G. Calcagni, M. Montobbio, and  G. Nardelli, \textit{A route to
nonlocal cosmology}, Phys. Rev. \textbf{D 76} (2007) 126001,
{\tt arXiv:0705.3043}\\
G. Calcagni and  G. Nardelli,\textit{ Tachyon solutions in boundary and
cubic string field theory}, Phys. Rev. \textbf{D 78} (2008) 126010, {\tt arXiv:0708.0366};\\
G. Calcagni, M. Montobbio, and  G. Nardelli, \textit{Localization of
nonlocal theories},
Phys. Lett. \textbf{B 662} (2008) 285--289, {\tt arXiv:0712.2237};\\
G. Calcagni and  G. Nardelli, \textit{Nonlocal instantons and solitons
in string models},  Phys. Lett. \textbf{B 669} (2008) 102--112, {\tt
arXiv:0802.4395}

 \bibitem{LJ-PR} L.V. Joukovskaya, {\it Dynamics in nonlocal cosmological models derived from string field theory}
  Phys. Rev. \textbf{D 76} (2007) 105007, {\tt arXiv:0707.1545};\\
   L.V. Joukovskaya, {\it Rolling tachyon in nonlocal cosmology}, L.V. Joukovskaya, AIP
Conf. Proc. \textbf{957}, \textit{Particles, Strings, and Cosmology},
eds. A. Rajantie, P.
Dauncey, C. Contaldi, H. Stoica, AIP, Melville, NY, 2007, pp. 325--328, {\tt arXiv:0710.0404};\\
L.V. Joukovskaya, {\it  Dynamics with Infinitely Many Time Derivatives
in Friedmann--Robertson--Walker Background
 and Rolling Tachyon},
  JHEP \textbf{0902} (2009) 045, {\tt arXiv:0807.2065}

\bibitem{noghosts}
 N. Barnaby and  N. Kamran,  \textit{Dynamics
with Infinitely Many Derivatives: The Initial Value Problem}, JHEP {\bf
0802} (2008) 008, {\tt arXiv:0709.3968}

\bibitem{AJV0711} I.Ya. Aref'eva, L.V. Joukovskaya, and S.Yu. Vernov,
 \textit{Dynamics in nonlocal linear models in the Friedmann--Robertson--Walker metric},
 J. Phys. A: Math. Theor. \textbf{41} (2008) 304003,
 {\tt arXiv:0711.1364}

\bibitem{Vernov:2008hd}
 S.Yu. Vernov, \textit{Exact Solutions in Nonlocal Linear Models}, Proc.
  the Int. Workshop SQS'07, (Dubna, Russia, July 30 - August 4, 2007),
  eds. E. Ivanov, S. Fedoruk, JINR, Dubna, Russia, 2008, pp. 118--121,
  {\tt arXiv:0802.3324}

\bibitem{MN}  D.J. Mulryne and N.J. Nunes,   \textit{Diffusing non-local inflation: Solving
the field equations as an initial value problem},
Phys. Rev. \textbf{D 78} (2008) 063519, {\tt arXiv:0805.0449};\\
D.J. Mulryne and N.J. Nunes, \textit{Non-linear non-local Cosmology},
AIP Conf. Proc. \textbf{1115} \textit{The Dark Side of the Universe},
eds. Sh.~Khalil, AIP, Melville, NY, 2009, pp.~329--334, {\tt
arXiv:0810.5471}

\bibitem{BarnabyKamran} N. Barnaby and
N. Kamran,  \textit{Dynamics with Infinitely Many Derivatives: Variable
Coefficient Equations}, JHEP {\bf 0812} (2008) 022, {\tt
arXiv:0809.4513}


\bibitem{KV}
A.S. Koshelev and S.Yu. Vernov, \textit{Cosmological perturbations in
SFT inspired non-local scalar field models}, {\tt arXiv:0903.5176}

\bibitem{CN}
G. Calcagni and G. Nardelli, \textit{Cosmological rolling solutions of
nonlocal theories}, Int. J. Mod. Phys. \textbf{D 19} (2010) 329--338, {\tt
arXiv:0904.4245}

\bibitem{Vernov2010} S.Yu. Vernov,  \textit{Localization of nonlocal cosmological models
with quadratic potentials in the case of double roots}, Class. Quant.
Grav. \textbf{27} (2010) 035006, {\tt arXiv:0907.0468}

\bibitem{Koshelev2009} A.S. Koshelev, \textit{SFT non-locality in cosmology:
solutions, perturbations and observational evidences}, {\tt
arXiv:0912.5457}

\bibitem{AV-NEC}
I.Ya.~Aref'eva and I.V.~Volovich, \textit{On the null energy condition
and cosmology}, Theor. Math. Phys. \textbf{155} (2008)  503--511 [Teor.
Mat. Fiz. \textbf{155} (2008) 3--12], {\tt hep-th/0612098}

\bibitem{RAS}
R.~Kallosh, J.U. Kang, A.~Linde, and V.~Mukhanov, {\it The New
Ekpyrotic Ghost}, JCAP {\bf 0804} (2008) 018, {\tt arXiv:0712.2040}

\bibitem{SW}
  S.~Weinberg,
  \textit{Effective Field Theory for Inflation},
  Phys. Rev.  \textbf{D 77} (2008) 123541,
  {\tt  arXiv:0804.4291};\\
  J.Z.~Simon,
  \textit{Higher derivative Lagrangians, non-locality, problems and
  solutions},
  Phys. Rev.  \textbf{D 41} (1990) 3720--3733

\bibitem{Creminelli0812} P. Creminelli, G. D'Amico, J. Norena, and  F.
Vernizzi, \textit{The Effective Theory of Quintessence: the $w<-1$ Side
Unveiled}, JCAP \textbf{0902} (2009) 018, {\tt arXiv:0811.0827}

\bibitem{data}
 A.G. Riess \textit{et al.}  [Supernova Search Team collaboration],
\textit{Type Ia Supernova Discoveries at $z>1$ From the Hubble Space
Telescope: Evidence for Past Deceleration and Constraints on Dark
Energy Evolution},  Astrophys. J. \textbf{607} (2004) 665--687, {\tt astro-ph/0402512};\\
 M. Tegmark \textit{et al.} [SDSS collaboration], \textit{The 3D power
spectrum of galaxies from the SDS},  Astroph.~J. \textbf{606} (2004) 702--740, {\tt astro-ph/0310725};\\
 P. Astier {\it et al.},  \textit{The Supernova Legacy Survey: Measurement
of $\Omega_M$, $\Omega_\Lambda$ and $w$ from the First Year Data Set},
Astron. Astrophys. \textbf{447} (2006) 31--48,
{\tt astro-ph/0510447};\\
Shirley Ho, Chr. M. Hirata, N.  Padmanabhan, U. Seljak, N. Bahcall,
\textit{Correlation of CMB with large-scale structure: I. ISW
Tomography and Cosmological Implications},
Phys. Rev. \textbf{D 78} (2008) 043519,   {\tt arXiv:0801.0642};\\
 W.M. Wood-Vasey {\it et al.} [ESSENCE Collaboration], \textit{Observational
Constraints on the Nature of the Dark Energy: First Cosmological
Results from the ESSENCE Supernova Survey},  Astrophys. J. \textbf{666}
(2007) 694--715, {\tt astro-ph/0701041};\\
D.~Baumann {\it et al.}  [CMBPol Study Team Collaboration],
  \textit{CMBPol Mission Concept Study: Probing Inflation with CMB Polarization},
  AIP Conf. Proc. \textbf{1141}, \textit{CMB Polarization workshop: theory and foregrounds: CMBPol
  mission concept}, eds. S. Dodelson, D. Baumann et al.,
  AIP, Melville, NY, 2009, pp. 10--120, {\tt arXiv:0811.3919};\\
E. Komatsu, J. Dunkley, M.R. Nolta, C.L. Bennett, B. Gold, G. Hinshaw,
N. Jarosik, D.~Larson, M. Limon, L. Page, D.N. Spergel, M. Halpern,
R.S. Hill, A. Kogut, S.S.~Meyer, G.S. Tucker, J.L. Weiland, E. Wollack,
E.L. Wright, \textit{Five-Year Wilkinson Microwave Anisotropy Probe
(WMAP) Observations: Cosmological Interpretation}, {Astrophys.\ J.\
Suppl.\ } \textbf{180} (2009) {330--376},
{\tt arXiv:0803.0547}; \\
M. Kilbinger, K. Benabed, J. Guy, P. Astier, I. Tereno, L. Fu, D.
Wraith, J. Coupon, Y.~Mellier, C. Balland, F.R. Bouchet, T. Hamana, D.
Hardin, H.J. McCracken, R.~Pain, N. Regnault, M. Schultheis, and H.
Yahagi, \textit{Dark energy constraints and correlations with
systematics from CFHTLS weak lensing, SNLS supernovae Ia and WMAP5},
Astron. Astrophys.  \textbf{497} (2009) 677--688, {\tt arXiv:0810.5129}


\bibitem{ZhangGui} Jingfei Zhang and Yuan-Xing Gui,
\textit{Reconstructing quintom from WMAP 5-year observations:
Generalized ghost condensate}, Commun. Theor. Phys., \textbf{54} (2010) 380--388, {\tt arXiv:0910.1200}

\bibitem{Starobinsky09} A. Shafieloo, V. Sahni, A.A. Starobinsky, \textit{Is cosmic acceleration slowing
down?}, Phys. Rev. \textbf{D 80} (2009) 101301, {\tt arXiv:0903.5141}

\bibitem{Quinmodrev1}
Yi-Fu Cai,  E.N. Saridakis, M.R. Setare, and Jun-Qing Xia,
\textit{Quintom Cosmology: theoretical implications and observations}, Phys. Rep. \textbf{493} (2010) 1--60,
{\tt arXiv:0909.2776};\\
 Hongsheng Zhang, \textit{ Crossing the
phantom divide},  {\tt arXiv:0909.3013}

\bibitem{string-cosmo} F. Quevedo, \textit{Lectures on string/brane cosmology},
Class. Quant. Grav. \textbf{19} (2002) 5721--5779, {\tt hep-th/0210292};\\
U.H. Danielsson, \textit{Lectures on string theory and cosmology},
Class. Quant. Grav. \textbf{22} (2005) S1-S40,  {\tt hep-th/0409274};\\
    M. Trodden and S.M. Carroll, \textit{TASI Lectures: Introduction to Cosmology}, {\tt astro-ph/0401547};\\
    A. Linde, \textit{Inflation and String Cosmology},
    J. Phys. Conf. Ser. \textbf{24} (2005) 151--160, {\tt hep-th/0503195};\\
C.P. Burgess, \textit{Strings, Branes and Cosmology: What can we hope to learn?}, {\tt hep-th/0606020};\\
        J.M.~Cline, \textit{String Cosmology},  {\tt hep-th/0612129};\\
  L. McAllister and  E. Silverstein, \textit{String Cosmology: A Review},
  Gen. Rel. Grav. \textbf{40} (2008) 565--605,  {\tt arXiv:0710.2951}

\bibitem{nonlocal}
 N. Arkani-Hamed,  S. Dimopoulos, G. Dvali, and  G. Gabadadze,
\textit{Nonlocal modification of gravity and the cosmological constant
problem}, {\tt hep-th/0209227};\\
 A.O. Barvinsky,  \textit{Nonlocal action for long-distance modifications
of gravity theory}, Phys. Lett. \textbf{B 572} (2003) 109--116,
{\tt hep-th/0304229};\\
S. Deser and  R.P. Woodard, \textit{Nonlocal Cosmology}, Phys. Rev. Lett.
\textbf{99} (2007) 111301, {\tt arXiv:0706.2151};\\
S. Nojiri and  S.D. Odintsov, \textit{Modified non-local-$F(R)$ gravity as
the key for the inflation and dark energy}, Phys. Lett. \textbf{B 659}
(2008) 821--826, {\tt arXiv:0708.0924};\\
S. Jhingan, S. Nojiri, S.D. Odintsov, M. Sami, I Thongkool, and  S. Zerbini,
\textit{Phantom and non-phantom dark energy: The cosmological relevance
of non-locally corrected gravity}, Phys. Lett. \textbf{B 663} (2008)
424--428, {\tt arXiv:0803.2613};\\
 S. Capozziello, E. Elizalde, Sh. Nojiri, and  S.D. Odintsov,  \textit{Accelerating
cosmologies from non-local higher-derivative gravity},
Phys. Lett. \textbf{B 671} (2009) 193--198, {\tt arXiv:0809.1535};\\
T.S. Koivisto, \textit{Newtonian limit of nonlocal cosmology},  Phys.
Rev. \textbf{D 78} (2008) 123505, {\tt arXiv:0807.3778};\\
S. Capozziello, E. Elizalde, Sh. Nojiri, and  S.D. Odintsov,
\textit{Accelerating cosmologies from non-local higher-derivative
gravity}, Phys. Lett. \textbf{B 671} (2009) 193--198,
{\tt arXiv:0809.1535}; \\
F.W. Hehl and  B. Mashhoon, \textit{A formal framework for a nonlocal
generalization of Einstein's theory of gravitation}, Phys. Rev.
\textbf{D 79} (2009) 064028, {\tt arXiv:0902.0560};\\
S. Nesseris and  A. Mazumdar, \textit{Newton's constant in $f(R,R_{\mu
\nu}R^{\mu \nu}$,$\Box R$) theories of gravity and constraints from
BBN}, Phys. Rev. \textbf{D 79} (2009) 104006, {\tt arXiv:0902.1185};\\
C. Deffayet and  R.P. Woodard, \textit{Reconstructing the Distortion
Function for Nonlocal Cosmology}, JCAP \textbf{0908} (2009) 023, {\tt arXiv:0904.0961};\\
 G. Cognola, E. Elizalde, S. Nojiri, S.D. Odintsov, and S. Zerbini,
\textit{One-loop effective action for non-local modified Gauss-Bonnet
gravity in de Sitter space}, Eur. Phys. J \textbf{C 64} (2009)
483--494, {\tt arXiv:0905.0543}

\bibitem{PaisU} A. Pais and  G.E. Uhlenbeck,
\textit{On Field Theories with Nonlocalized Action},
 Phys. Rev. \textbf{79} (1950) 145--165

\bibitem{STinspired}
 D.A. Eliezer and  R.P. Woodard, \textit{The Problem of
Nonlocality in
String Theory}, Nucl. Phys. \textbf{B 325} (1989) 389--469;\\
J. Llosa and J. Vives, \textit{Hamiltonian formalism for nonlocal
Lagrangians},
J. Math. Phys. \textbf{35} (1994) 2856--2877; \\
R.P. Woodard, \textit{A Canonical Formalism For Lagrangians With
Nonlocality Of Finite Extent}, Phys. Rev. \textbf{A 62} (2000) 052105;\\
 N. Moeller and  B. Zwiebach, \textit{Dynamics with
Infinitely Many Time Derivatives and Rolling Tachyons},  JHEP {\bf
0210} (2002)
034, {\tt hep-th/0207107};\\
V.S. Vladimirov and  Ya.I. Volovich, \textit{Nonlinear Dynamics
Equation in p-Adic String Theory}, Theor. Math. Phys. {\bf 138} (2004)
297--309 [Teor. Mat. Fiz., {\bf 138} (2004)
355--368], {\tt math-ph/0306018};\\
A. Sen, {\it Tachyon Dynamics in Open String Theory},
 Int. J. Mod. Phys. \textbf{A 20} (2005) 5513--5656,
{\tt hep-th/0410103}\\
V.S. Vladimirov, {\it On the equation of the $p$-adic open string
for the scalar tachyon field},  {\tt math-ph/0507018}; \\
V. Forini, G. Grignani, and  G. Nardelli,
 {\it A new rolling tachyon solution of cubic string field theory},
 JHEP \textbf{0503} (2005) 079, {\tt hep-th/0502151};\\
L.V.~Joukovskaya, {\it Iteration method of solving nonlinear integral
equations describing rolling solutions in string theories}, Theor.
Math. Phys. {\bf 146} (2006)  335--342
[Teor. Mat. Fiz. {\bf 146} (2006) 402--409],  {\tt arXiv:0708.0642};\\
B. Dragovich, \textit{Zeta Nonlocal Scalar Fields}, Theor. Math. Phys.
\textbf{157} (2008) 1671--1677
 [Teor. Mat. Fiz. {\bf 157} (2008) 364--372], {\tt arXiv:0804.4114};\\
G. Calcagni and G. Nardelli, \textit{Kinks of open superstring field
theory}, Nucl. Phys. \textbf{B 823} (2009) 234--253, {\tt
arXiv:0904.3744}

\bibitem{Yang} H. Yang,
{\it Stress tensors in p-adic string theory and truncated OSFT}, JHEP
\textbf{0211} (2002) 007, {\tt hep-th/0209197}


\bibitem{AJK}  I.Ya. Aref'eva, L.V. Joukovskaya, and  A.S. Koshelev,
\textit{Time evolution in superstring field theory on nonBPS brane. 1.
Rolling tachyon and energy momentum conservation},
 JHEP \textbf{0309} (2003) 012, {\tt hep-th/0301137}


\bibitem{davis}
H.T. Davis, \textit{The Laplace differential equation of infinite
order}, Anal. Math. {\bf 2} 32 (1931) no. 4, 686--714;\\
 H.T. Davis, \textit{The Theory of Linear Operators from the
Standpoint of Differential Equations of Infinite Order}, Indiana, the
Principia Press, 1936

\bibitem{carmi}
R.D. Carmichael, \textit{Linear differential equations of infinite
order}, Bull. Amer. Math. Soc.  \textbf{42} (1936)  193--218;\\
 L. Carleson, \textit{On infinite differential equations with constant
coefficients. I}, Math. Scand.  \textbf{1} (1953) 31--38


\bibitem{TCSFT} A.Sen and  B. Zwiebach, {\it Tachyon condensation in string field theory},
  \textit{JHEP} \textbf{0003} (2000) 002, {\tt hep-th/9912249}

 \bibitem{AK} I.Ya.~Aref'eva and  A.S.~Koshelev,
 {\it Cosmic Acceleration and Crossing of $w=-1$ barrier
 from Cubic Superstring Field Theory},  \textit{JHEP} \textbf{0702}
 (2007) 041, {\tt hep-th/0605085}

 \bibitem{ABJV0903}
I.Ya. Aref'eva, N.V. Bulatov, L.V. Joukovskaya, and  S.Yu. Vernov,
\textit{Null Energy Condition Violation and Classical Stability in the
Bianchi I Metric},   Phys. Rev. \textbf{D 80} (2009) 083532, {\tt
arXiv:0903.5264}

\bibitem{Fisher} I.Z. Fisher,
\textit{Scalar mesostatic field with regard for gravitational effects},
Zh. Eksp. Teor. Fiz. \textbf{18} (1948) 636--640, {\tt gr-qc/9911008}

\bibitem{AboutFisher1}
    K.A. Bronnikov, M.S. Chernakova, J.C. Fabris, N. Pinto-Neto, and  M.E.
    Rodrigues, \textit{Cold black holes and conformal continuations},
    Int. J. Mod. Phys. \textbf{D 17} (2008) 25--42,
    {\tt gr-qc/0609084}

\bibitem{AboutFisher2}
Sh. Abdolrahimi and  A.A. Shoom, \textit{Analysis of the Fisher solution},
Phys. Rev. \textbf{D 81} (2010) 024035, {\tt arXiv:0911.5380} }
\end{thebibliography}
\end{document}